\documentclass[useAMS,usenatbib]{mn2e}
\input amssym.tex
\usepackage{times}
\usepackage{amstext}
\usepackage{bm}
\usepackage{url}
\usepackage{graphicx}
\usepackage{hyperref}

\mathchardef\mhyphen="2D

\newcommand{\bssC}{\mathbf{\mathsf{C}}}
\newcommand{\Cinit}{\mathbf{\mathsf{C^{init}}}}

\newcommand{\bssG}{\mathbf{\mathsf{G}}}
\newcommand{\bssF}{\mathbf{\mathsf{F}}}
\newcommand{\bq}{\textbf{q}}

\newcommand{\bx}{\textbf{x}}

\newcommand{\PDF}{\mathcal{P}}

\newcommand{\avg}[1]{\left\langle{#1}\right\rangle}

\newcommand{\hmpc}{\,$h^{-1}$\,Mpc}

\newcommand{\hmpcnosp}{$h^{-1}$\,Mpc}
\newcommand{\ihmpc}{\,$h$\,Mpc$^{-1}$}

\newcommand{\LCDM}{$\Lambda$CDM}
\newcommand{\pd}{$P_\delta$}
\newcommand{\precip}{$P_{1/(1+\delta)}$}

\newcommand{\pln}{$P_{\ln(1+\delta)}$}
\newcommand{\pg}{$P_{{\rm Gauss}(\delta)}$}
\newcommand{\dg}{${\rm Gauss}(\delta)$}

\newcommand{\Var}{{\rm Var}}

\newcommand{\bPsi}{\mbox{\boldmath $\Psi$}}

\newcommand{\drec}{$\frac{1}{1+\delta}$}

\newcommand{\gij}{G_{ij}}
\newcommand{\gdij}{G^{\delta}_{ij}}

\newcommand{\ggij}{G^{{\rm Gauss}(\delta)}_{ij}}
\newcommand{\grecij}{G^{1/(1+\delta)}_{ij}}
\newcommand{\pinit}{P^{\rm init}}

\chardef\til=`\~

\begin{document}

\title[Response of power spectra  to initial spikes]{Ringing the initial Universe: the response of overdensity and transformed-density power spectra to initial spikes}

\author[Mark C.\ Neyrinck, Lin Forrest Yang]
{Mark C.\ Neyrinck$^1$ and Lin Forrest Yang$^1$\\
$^{1}$Department of Physics and Astronomy, The Johns Hopkins University, Baltimore, MD 21218, USA}


\maketitle

\begin{abstract}
  We present an experiment in which we `ring' a set of cosmological $N$-body-simulation initial conditions, placing spikes in the initial power spectrum at different wavenumber bins.  We then measure where these spikes end up in the final conditions.  In the usual overdensity power spectrum, most sensitive to contracting and collapsing dense regions, initial power on slightly non-linear scales ($k\sim 0.3$\ihmpc) smears to smaller scales, coming to dominate the initial power once there.   Log-density and Gaussianized-density power spectra, sensitive to low-density (expanding) and high-density regions, respond differently: initial spikes spread symmetrically in scale, both upward and downward.  In fact, in the power spectrum of $1/(1+\delta)$, spikes migrate to larger scales, showing the magnifying effect of voids on small-scale modes. These power spectra show much greater sensitivity to small-scale initial features.  We also test the difference between an approximation of the Ly-$\alpha$ flux field, and its Gaussianized form, and give a toy model that qualitatively explains the symmetric power spreading in Gaussianized-density power spectra.  Also, we discuss how to use this framework to estimate power-spectrum covariance matrices.   This can be used to track the fate of information in the Universe, that takes the form of initial degrees of freedom, one random spike per initial mode.
\end{abstract}

\begin {keywords}
  large-scale structure of Universe -- cosmology: theory
\end {keywords}

\section{Introduction}
Fluctuations in the present-epoch cosmic density field are rich in cosmological information.  However, even for the theoretically straightforward real-space dark-matter field, the relation between the initial and final fluctuations is obscure on small scales.  In the usual overdensity $\delta=\rho/\bar{\rho}-1$, the shape of the power spectrum departs substantially from linear theory on small scales.  A further obstacle to inferring initial information is the substantial covariance which arises in the $\delta$ power spectrum on small scales \citep{MeiksinWhite1998}.  This covariance greatly reduces the Fisher information, i.e.\ increases error bars on cosmological parameters observable in principle \citep{RimesHamilton2006, NeyrinckEtal2006, TakahashiEtal2009, KiesslingEtal2011}.

Fortunately, much of this apparently lost Fisher information can recovered using a local 1-point probability density function (PDF) Gaussianizing transform, such as a logarithm or rank-order-Gaussianization \citep{NeyrinckEtal2009,SeoEtal2011,Neyrinck2011b,YuEtal2011}.   These Fisher analyses used measurements of final-conditions covariances in the power spectrum, which are related to the independent degrees of freedom resident there, but not directly to the initial degrees of freedom.

In this Letter, we explicitly track these initial degrees of freedom, in an $N$-body experiment in which we observe the results of `ringing' the initial density field with initial power-spectrum spikes at different wavenumbers.  This quantifies the `memory' of initial conditions (ICs) in the final field, but differently than the propagator \citep[][CS06]{CrocceScoccimarro2006}, which can be thought of as a cross-correlation of a mode's amplitude and phase in the initial and final conditions.   Information in an initial mode gets deposited in larger or smaller-scale modes, but does not disappear, as a naive interpretation of a vanishing propagator might suggest.

We measure the response in the power spectra of a few transformed fields: $e^{-(1+\delta)}$; the log-density $\ln(1+\delta)$; the rank-order-Gaussianized density \dg; and the reciprocal-density \drec.  The `${\rm Gauss}$' function is an increasing function, depending on the PDF of its argument, whose result has a Gaussian PDF.  Explicitly,  ${\rm Gauss}(\delta) =\sqrt{2}\sigma {\rm erf}^{-1}(2f_{<\delta}-1+1/N)$, where $f_{<\delta}$ is the fraction of cells less-dense than $\delta$, $\sigma$ is the standard deviation of the Gaussian that $\delta$'s PDF is mapped onto, and $N$ is the number of cells.  Below, we set $\sigma = \sqrt{\Var[\ln(1+\delta)]}$.  So \dg\ is essentially $\ln(1+\delta)$ with its skewness removed.

These transformations increasingly emphasize low-density regions, where the initial fluctuations are the most pristine.  We expect the different effective weightings provided by the transformations to affect the way power migrates in scale as structure forms. \citet{McCullaghEtal2013} showed analytically using the Zel'dovich approximation \citep[ZA, ][]{Zeldovich1970} that the baryon acoustic (BAO) peak location in the correlation function changes if a transform is applied to the field, based on the density regimes that the transform emphasizes.  In $\delta$, $\ln(1+\delta)$, and $\frac{1}{1+\delta}$, the BAO peak location is slightly biased inward, nearly unbiased, and biased outward compared to linear theory.  The present Letter carries these results to general power-spectrum features.

We investigate these particular transformations for various reasons.  In a lognormal \citep{ColesJones1991}, or nearly lognormal field, two-point statistics of the log-density capture a much greater fraction of the Fisher information in the field than those of $\delta$ \citep{NeyrinckEtal2009,Carron2011,CarronNeyrinck2012}.  Forcing the one-point moments to zero, as Gaussianization does, generally decreases higher-order multi-point correlations as well, and allows two-point statistics to more effectively capture the information in a field.  Also, the power spectrum of the log-density and Gaussianized density have an intriguingly nearly-linear shape, which this study addresses.  \drec\ is of some theoretical interest, being a simple quantity in a Lagrangian framework (just the Jacobian of the deformation tensor); also, it is rather insensitive to multistreaming, which only further suppresses high-density regions.  We investigate $e^{-(1+\delta)}$ because of its relevance to Ly-$\alpha$-forest measurements; from a dark-matter simulation, it gives the flux, in the (poor) approximation that neutral hydrogen follows dark matter.  It is of interest in the study of density transforms, since Gaussianization \citep{Weinberg1992} was tried in Ly-$\alpha$-forest studies \citep{CroftEtal1998}, but later works skipped that step \citep[e.g.\ ][]{CroftEtal2002}.  Without noise, ${\rm Gauss}(\delta)=-{\rm Gauss}(e^{-(1+\delta)})$, so comparing the results of these two transformations is a test of Gaussianization's usefulness in Ly-$\alpha$ studies.

In Section \ref{sec:methods}, we describe the simulation suites and other techniques.  In Section \ref{sec:results}, we give our main results.  In Section \ref{sec:model}, we explain a crude toy model that qualitatively captures the migration of power in Gaussianized-density power spectra.  In Section \ref{sec:info}, we discuss estimating practical power-spectrum covariance matrices and Fisher information from these results.

\section{Methods}
\label{sec:methods}

\begin{figure}
  \begin{center}
    \includegraphics[width=\columnwidth]{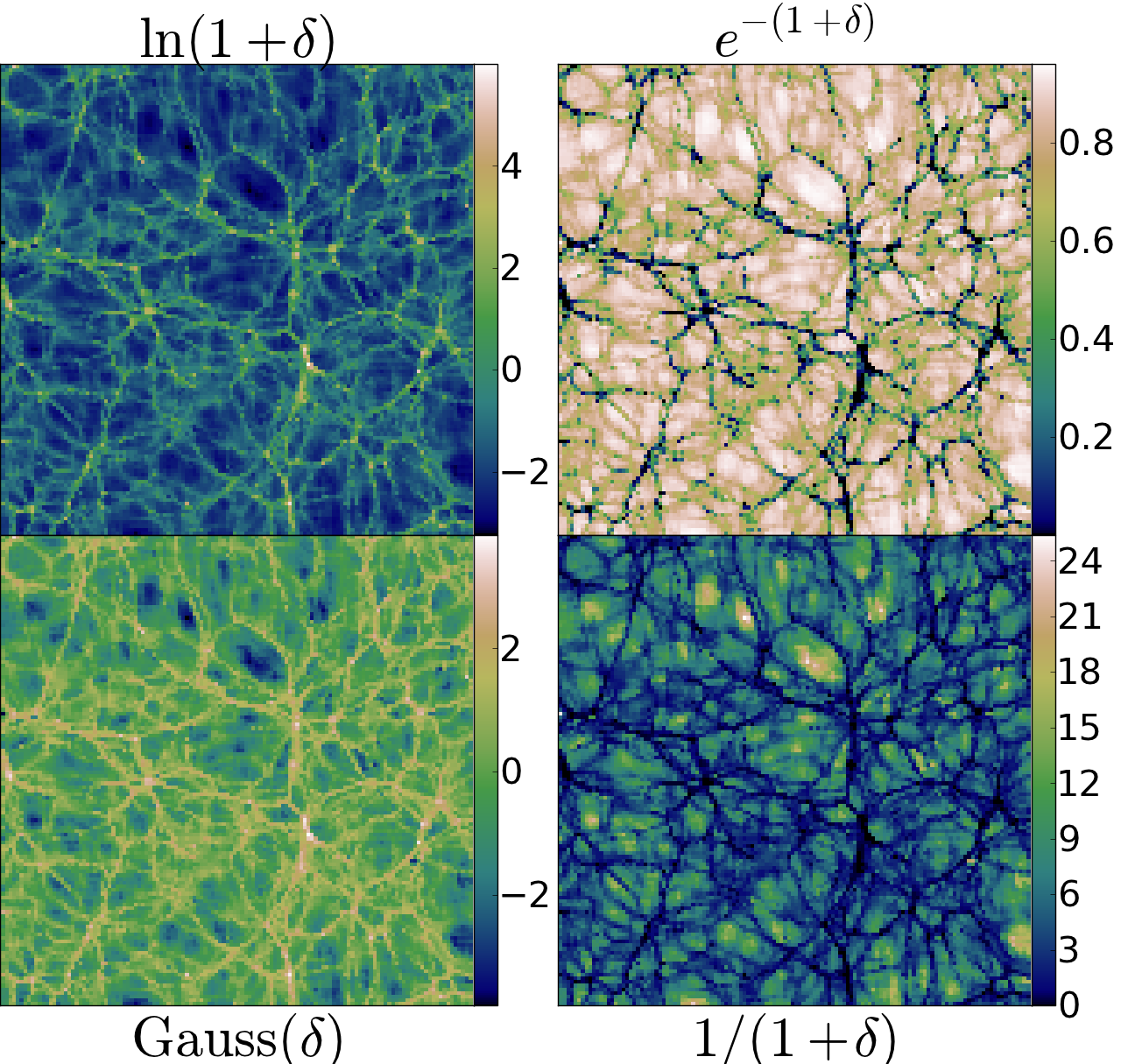}
  \end{center}  
  \caption{Plots of the transformed density-field slices from simulations used.  The width of each panel is 128\hmpc, half the box size, and the pixel size is 1\hmpc.  The Lagrangian-tessellation density estimate allows accurate density estimates deep within voids, with suppressed particle-discreteness effects compared to other methods.}
  \label{fig:transformpanels}
\end{figure}

We ran three ensembles of 24 simulations: one `control' simulation without initial power-spectrum spikes, and 23 simulations with spikes, each in a different power-spectrum bin.  Previous numerical experiments have been run in which some Fourier modes are kept fixed, and others are changed \citep[e.g.\ ][]{ShandarinMelott1990, LittleEtal1991, SuhhonekoEtal2011, AragonCalvo2012}.  Our simulations, run using {\scshape Gadget}2 \citep{Springel2005}, each have $256^3$ particles in a 256\hmpc\ box, and ICs generated using the ZA at redshift $z=127$.  The simulations used vanilla \LCDM\ cosmological parameters: $(h, \Omega_b, \Omega_{\rm cdm}, \Omega_\Lambda, \sigma_8, n_s)=(0.73, 0.045, 0.205, 0.75, 0.8, 1)$.  Each spike was inserted by multiplying the mode amplitudes in the corresponding bin by $\sqrt{2}$, holding all other aspects of the ICs fixed.  As usual, each initial mode's phase is random and uncorrelated to other phases.  However, to avoid possible effects from spikes that arise randomly from cosmic variance, the amplitude of each initial mode at $k$ is set to exactly $\sqrt{P(k)}$.  So, the only difference between the three ensembles is in their sets of random phases.  Because Fourier amplitudes are not drawn from a Rayleigh distribution as usual, some higher-order statistics within each box may be suppressed, but the simulations are suitable for their intended purpose, to track power migration.

On the scales investigated, $P_\delta\gg 1/n$, the $P_\delta$ shot noise.  For the $\delta$ analysis, we use a cloud-in-cell (CIC) density assignment on a $256^3$ grid.  The largest wavenumber plotted is that where power spectra estimated with a nearest-grid-point (NGP) method departs noticeably from the CIC power spectra.

However, for $\ln(1+\delta)$, \dg, and especially \drec, care is required in density estimation because of sensitivity to low-density regions.   The results also depend somewhat on the cell size to which the transforms are applied; here, we use 1\hmpc\ cells.  We use what we call the Lagrangian Tessellation Field Estimator \citep[LTFE,\ ][]{AbelEtal2012,ShandarinEtal2012}, referencing the Delaunay and Voronoi Tessellation Field Estimators \citep[DTFE, VTFE, e.g.\ ][]{SchaapVdW2000}, which use Eulerian tessellations.    Fig.\ \ref{fig:transformpanels} shows transformed-density slices measured this way, from our simulations.  The Lagrangian tessellation has a more physical meaning than the Eulerian tessellations, although it has the drawback that it requires knowledge of the ICs, and thus cannot directly be applied to observations.  In the LTFE, particles are treated not as mass blobs, but as vertices on a dark-matter sheet \citep[see also][]{FalckEtal2012,Neyrinck2012}.  For most purposes, and especially for low densities, this approximation is better than treating them as mass blobs, since a physical dark-matter particle is dozens of orders of magnitude lighter than a simulation `particle.'  Lagrangian space is tessellated into tetrahedra according to the initial cubic particle lattice.  Matter is uniformly deposited into tetrahedra, which often overlap in high-density regions where streams cross.  No pixel is empty, since for each position, there is at least one tetrahedron that gets stretched across it.  This method reveals fluctuations in low-density regions that would be difficult to see using a more naive density-estimation method.  Much higher mass resolution would be necessary to measure \precip\ using CIC.

Ideally, the LTFE would be computed in our cubic cells by computing the intersections of tetrahedra with each cubic cell.  However, for speed, we instead sample the density on a cubic lattice of points.  For each tetrahedron enclosing a given lattice point, we add densities inversely proportional to the tetrahedron's volume.  To reduce the noise from this point-sampling, each grid point comes from an eightfold density super-sampling; the density is measured on a $512^3$ grid, which we then average down to a $256^3$ grid.  Even so, the process requires much more computation than e.g.\ CIC, so we implemented a fast GPU code employing CUDA technology.

\section{Results}
\label{sec:results}
\begin{figure}
  \begin{center}
    \includegraphics[width=\columnwidth]{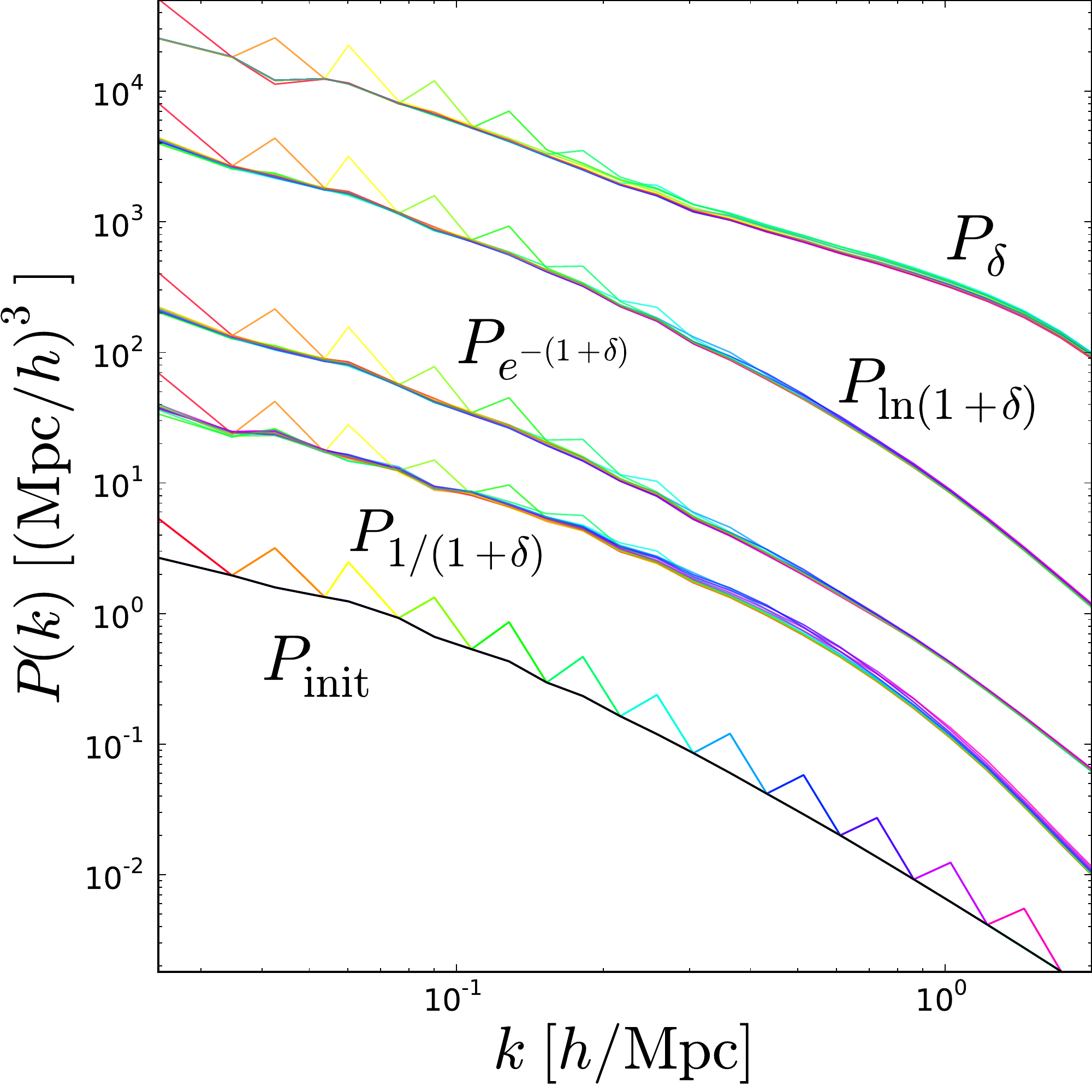}
  \end{center}  
  \caption{From top to bottom, $z=0$ power spectra from initial-spike simulations of various transformed-density power spectra, along with the spiked initial conditions at $z=127$.   Power spectra are rainbow-colored according to the wavelength of their initial spike, from short (violet) to long (red).  $P_{1/(1+\delta)}$ is divided by $10^3$ for clarity.
  $P_{{\rm Gauss}(\delta)}$, investigated below, is omitted because it is almost indistinguishable from \pln.}
  \label{fig:rawpowers}
\end{figure}

Fig.\ \ref{fig:rawpowers} shows various power spectra at $z=0$ from one ensemble of spiked simulations.   Also shown is the initial density field used to generate the ICs, $P_{\rm init}$.  At large scales, initial spikes are preserved in each power spectrum.  But on small scales, each power spectrum behaves differently.  At $k\sim 1$\ihmpc, the initial scales dominating \pd\ correspond to the green color (larger scales), while the initial scales dominating \precip\ correspond to violet (smaller scales).

\begin{figure}
  \begin{center}
    \includegraphics[width=\columnwidth]{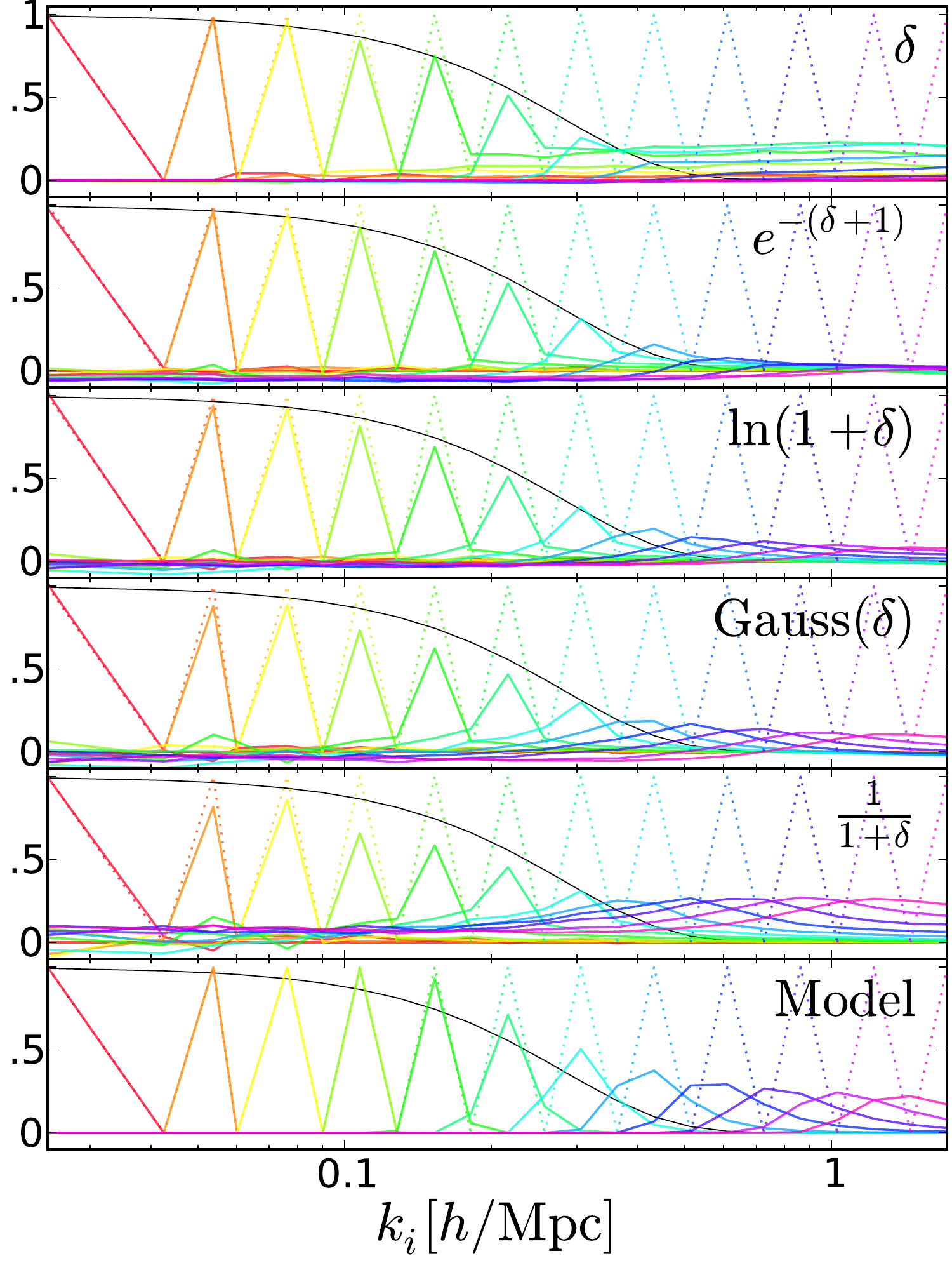}
  \end{center}  
  \caption{Plots of the matrix $\gij$, defined in Eq.\ (\ref{eqn:gijdef}), showing the response of final-conditions power spectra to initial spikes.   The spikes in the initial conditions are shown with dotted curves, rainbow-colored from red to violet going from low to high frequency.   Corresponding power spectra of final-conditions simulations appear as solid curves.  From top to bottom, the final-conditions resting place of a moderate-scale (e.g.\ green) spike moves from small to large scales, as each transformation increasingly emphasizes underdense regions.   For clarity, power spectra from only odd-numbered spikes are shown.  The `model' is a toy model of power spreading based on a local spherical collapse or expansion of volume elements, given in Eq.\ (\ref{eqn:scs}).  The black curves show the density propagator (CS06).
  	}
  \label{fig:nbody_spikelines}
\end{figure}

\begin{figure}
  \begin{center}
    \includegraphics[width=\columnwidth]{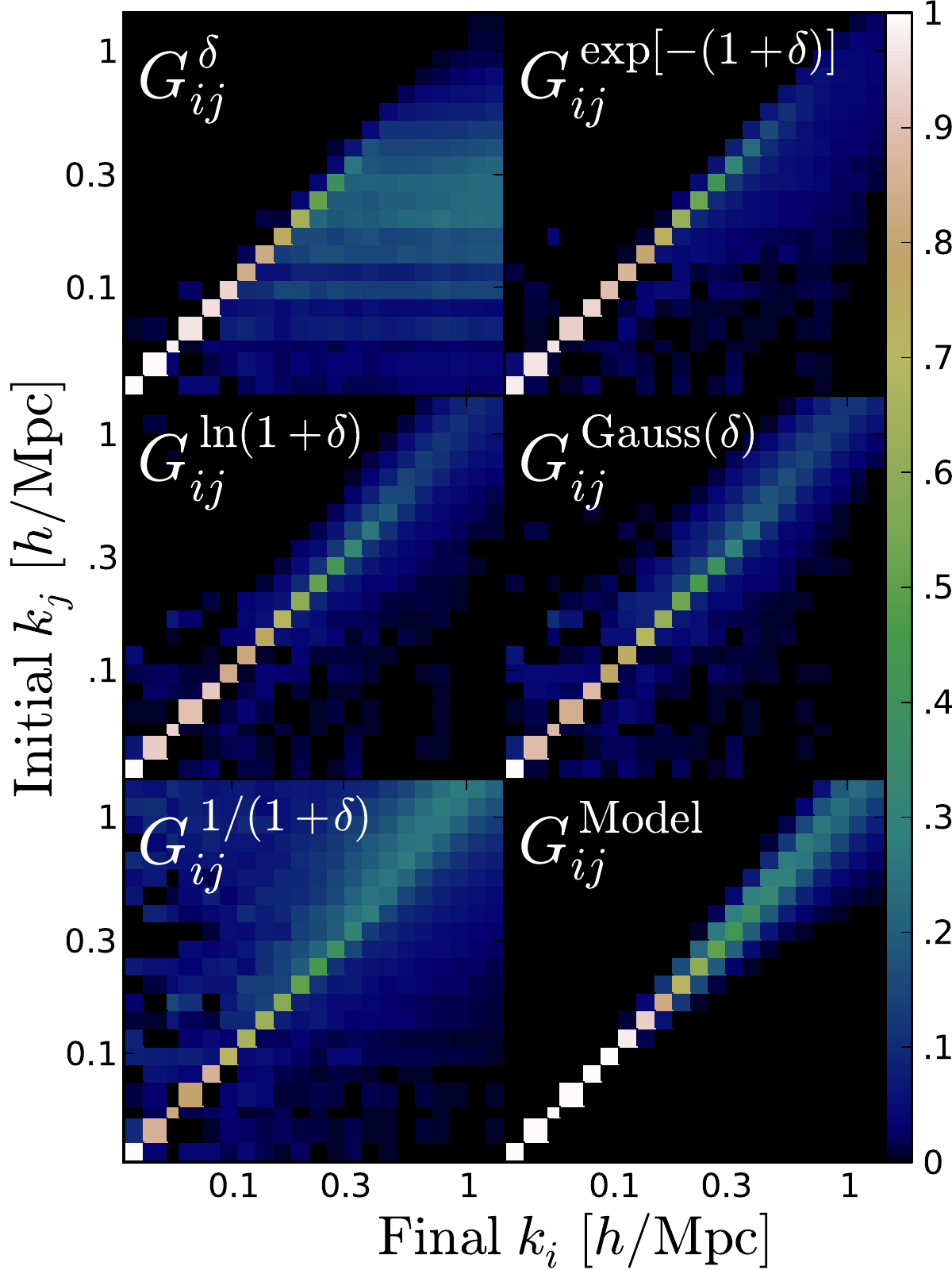}
  \end{center}  
  \caption{$G_{ij}$'s as in Fig.\ \ref{fig:nbody_spikelines}, shown in matrix form.}
  \label{fig:nbody_spikematrix}
\end{figure}

Figs.\ \ref{fig:nbody_spikelines} and \ref{fig:nbody_spikematrix} show the linear-response matrix 
\begin{equation}
\gij\equiv\frac{\partial \ln P(k_i)}{\partial \ln P^{\rm init}(k_j)}.
\label{eqn:gijdef}
\end{equation}
We use three ensembles of simulations to lower the realization-to-realization variance in the measured $\gij$, which is typically at the $\sim10\%$ level, depending on scale.  Fixing all mode amplitudes to their ensemble-mean values reduces this variance, but differences in random phases still produce fluctuations.  The reported value of each $G_{ij}$ matrix element is the median among the three ensembles.

In $\delta$, the migration of power is qualitatively as in \citet{HamiltonEtal1991} and \citet{PeacockDodds1996}: power moves from large to small scales.  However, a fixed initial scale does not migrate to a fixed non-linear scale.   Translinear ($k\sim 0.3$\ihmpc) initial scales end up smeared over a wide range of smaller scales.   The result is that small ($k\sim 1$\ihmpc) final scales are dominated by translinear initial modes, and are insensitive to initial power at their own comoving wavenumber.   Indeed, initial power inserted at $k\sim 1$\ihmpc\ hardly affects \pd\ over the scales measured, although it might show up at smaller scales in higher-resolution simulations.  Black curves show an approximation of the $z=0$ density propagator, a Gaussian with $\sigma_k=0.2$\ihmpc (CS06).  The propagator is a cross-correlation of modes in the initial and final conditions, sensitive to both amplitudes and phases, and, as expected, it follows $\gdij$'s diagonal quite well.

In $\ln(1+\delta)$ and \dg, the behavior is different.  Power spreads out, moving not just from large to small scales, but vice-versa, rather symmetrically.  This makes sense: \pd\ is mostly sensitive to high-density regions, where fluctuations have contracted.  \pln\ and \pg, on the other hand, are sensitive to dense regions, but also to low-density regions, where fluctuations have expanded.  In fact, initial peaks migrate to a bit larger scales than they were initially (visible as slight upturns in Fig.\ \ref{fig:nbody_spikematrix}).  There are a couple of possible reasons for this upturn: underdense regions, even if equal by mass, dominate by volume.  Also, in overdense regions, many fluctuations completely collapse, perhaps leaving no clear signal in the power spectrum.  In underdense final regions, however, almost all fluctuations remain, stretched-out compared to the ICs.  Indeed, in the \drec\ field, in which underdense regions receive almost all weight, the high value of $\grecij$ above the diagonal indicates this comoving expansion of fluctuations.  This explicitly confirms the magnifying effect \citep[][AS13]{AragonCalvoSzalay2013} that voids have on initially small scales.

Going from $\delta$ down to \drec\ in Fig.\ \ref{fig:nbody_spikelines}, the sensitivity to small-scale initial power increases substantially.  In the transformed fields, although each individual final mode still retains little memory of its initial phase and amplitude, as quantified by the propagator \citep{WangEtal2011}, small-scale spikes do leave behind obvious bumps at approximately their initial scale.  There is a tradeoff on large scales, though: diagonal entries decease slightly, and off-diagonal terms fluctuate increasingly from zero.  Large-scale modes, exceeding the scales of typical displacements, are fixed in \pd\ by mass conservation, but after a transformation, this is no longer assured, so some fluctuations in large-scale mode amplitudes occur.  However, there seems to be no systematic trend in these fluctuations; averaging over many realizations of phases, it seems that these off-diagonal elements would average to zero. 

\section{Toy model for power spreading}
\label{sec:model}
A toy model of $\ggij$ based on Lagrangian dynamics captures its behavior qualitatively.  The model, which does not explicitly consider the Gaussianization process, is that initial fluctuations expand or contract according to the local density, and that the final \pg\ manages to pick up every fluctuation at the scales where it has expanded or contracted in final conditions.  In reality, some initial fluctuations collapse, or otherwise escape detection by \pg, but the approximation here is that they do not.

Consider fluctuations imprinted on initial pixels of equal Lagrangian size that expand or contract according to the local density.  In the ZA, $\nabla_q\cdot \bPsi = -\delta_{\rm lin}$, where $\bPsi = \bx-\bq$ is the Lagrangian displacement field, and $\delta_{\rm lin}$ is the linearly extrapolated initial density.  Approximating each pixel's expansion or contraction as isotropic, with Lagrangian displacement-divergence `stretching' parameter $\psi\equiv \nabla_q\cdot\bPsi$, each of its 3 dimensions will scale by a factor $1+\psi/3$.  On logarithmic plots such as Figs.\ \ref{fig:nbody_spikelines} and \ref{fig:nbody_spikematrix}, a fluctuation occupying pixels that stretch in such a way gets shifted by $s\equiv -\ln (1+\psi/3)$, the minus sign because of the reciprocal from working in Fourier space. 

 A Gaussian PDF of $\psi$, with variance $\sigma$, can be transformed into a mass-weighted (Lagrangian volume) PDF of $s$ in the ZA:
 \begin{equation}
 	\PDF(s) = \PDF(\psi)\left|\frac{d\psi}{ds}\right| = \frac{3 \exp\left[-(3e^{-s}-1)^2/(2\sigma^2)-s\right]}{\sqrt{2\pi\sigma^2}},
	\label{eqn:zs}
\end{equation}
where $\PDF$ denotes a probability distribution.

Alternatively, a spherical-collapse approximation can be used for the behavior of $\psi$, where $\psi_{\rm z}$ is its ZA value  \citep{MohayaeeEtal2006,Neyrinck2013}, 
\begin{equation}
	\psi_{\rm sc} = 3\left[\left(1+\frac{2}{3}\psi_{\rm z}\right)^{1/2}-1\right].
\end{equation}
This uses an approximation for the evolution of a volume element found by \citet{Bernardeau1994}, valid in the limit of low matter density.  It gives the following distribution of $s$:
\begin{equation}
  \PDF(s) = \frac{(3/2) \exp\left[-(3/2)(e^{-2 s}-1)^2/(2\sigma^2)-2s\right]}{\sqrt{2\pi\sigma^2}}
  \label{eqn:scs}
\end{equation}
Both Eqs.\ (\ref{eqn:zs}) and (\ref{eqn:scs}) can be approximated by a Gaussian of dispersion $\sigma/3$ for small $\sigma$.

The final panels of Figs.\ \ref{fig:nbody_spikelines} and \ref{fig:nbody_spikematrix} show $G_{ij}$ in this model.
The assumption is that Eq.\ (\ref{eqn:scs}) gives the shapes of the curves into which spikes broaden in Fig.\ \ref{fig:nbody_spikelines}.  Each $j$ row is normalized so that all initial fluctuations contribute to some final wavenumber, i.e.\ so that for all $j$, $\Sigma_i G_{ij}^{\rm Model} = 1$.   Physically, this would describe $G_{ij}$ for a density variable that manages to capture all Lagrangian volumes, both expanding and contracting.  It is not surprising that $G_{ij}^{\rm Model}$ has higher amplitude on small scales than any measured $G_{ij}$, since some fluctuations doubtless evade notice by any final power spectrum, i.e.\ $\Sigma_i G_{ij} < 1$ generally.  We estimate $\sigma^2$ in Eq.\ (\ref{eqn:scs}) as the variance in top-hat spheres of radius $2\pi/k_j$ in linear theory.  This model is quite naive; for instance, it assumes an equal-Lagrangian-volume (i.e.\ mass) weighting, not an Eulerian equal-volume-weighting.  Still, the model captures the qualitative behavior of $\ggij$, although it may underestimate the variance at low $k_j$, and overestimates $\ggij$ along the diagonal.

\section{Covariance Matrices and Information}
\label{sec:info}
$G_{ij}$ can be used to estimate power-spectrum covariances, as well.  Usually, covariance matrices are measured directly from ensembles of final density fields, and it is interesting to compare this approach with the result of directly tracking initial-conditions degrees of freedom.  A linear model of a fluctuation away from the mean in the final-conditions $\ln P_i$ (investigated instead of $P_i$ for algebraic simplicity) is, summing over repeated indices,
\begin{equation}
  \Delta \ln P_i = \frac{\partial \ln P_i}{\partial \ln \pinit_j} \Delta \ln \pinit_j = \gij \Delta \ln \pinit_j,
  \label{eqn:deltalin}
\end{equation}
giving
\begin{equation}
  C_{ij} = \avg{\Delta \ln P_i \Delta \ln P_j} = \avg{G_{ik} \Delta \ln \pinit_k G_{jl} \Delta \ln \pinit_l}.
  \label{eqn:deltalin2}
\end{equation}
Since the Gaussian $C^{\rm init}_{ij} \equiv \avg{\Delta \ln \pinit_i \Delta \ln \pinit_j} = 2\delta^{\rm K}_{ij}/N_i $, where $N_i$ is the number of modes in bin $i$,
\begin{equation}
 \bssC= \bssG \Cinit \bssG^{\top}.
\end{equation}

Now, suppose we want to estimate all $\ln \pinit_i$ from the final power spectrum.  The Fisher matrix $F_{ij}$ to use to predict constraints on this initial power in bins $i$ and $j$ would be
\begin{equation}
 	F_{ij}=\frac{\partial \ln P_k}{\partial \ln P^{\rm init}_i}(\bssC^{-1})_{kl}\frac{\partial \ln P_l}{\partial \ln P^{\rm init}_j}=G_{ki}(\bssC^{-1})_{kl} G_{lj},
 \end{equation}
 or
\begin{equation}
 	\bssF=\bssG^\top(\bssG \Cinit \bssG^\top)^{-1}\bssG = (\Cinit)^{-1}.
 \end{equation}
So the covariance matrix of parameters that consist of the initial power spectrum in bins is just the initial, Gaussian power-spectrum covariance matrix.  This suggests no Fisher-information loss!

However, this calculation assumes that the final power spectrum is an entirely deterministic, invertible, linear transformation of the initial power spectrum, with no sources of noise.  This is not the case; it neglects at least a couple of things: the non-linear coupling of pairs of power-spectrum spikes, and realization-to-realization fluctuations in the $G_{ij}$ matrix, which can depend on both the power spectrum itself, and also on mode correlations (present even in a Gaussian field) that affect the halo mass function, which substantially affects translinear-scale power, at least in the halo model \citep{NeyrinckEtal2006}.

One neglected factor that can be investigated in a straightforward extension of the present framework is the non-linear coupling of spike pairs.  But this would involve a simulation for each pair of wavenumber bins, i.e.\ with 23 bins, $23\times 22=506$ additional simulations.  We plan to run this brute-force ensemble in future work.  For now, we compare the covariance matrix from Eq.\ (\ref{eqn:deltalin}) to one estimated otherwise.

\begin{figure}
  \begin{center}
    \includegraphics[width=\columnwidth]{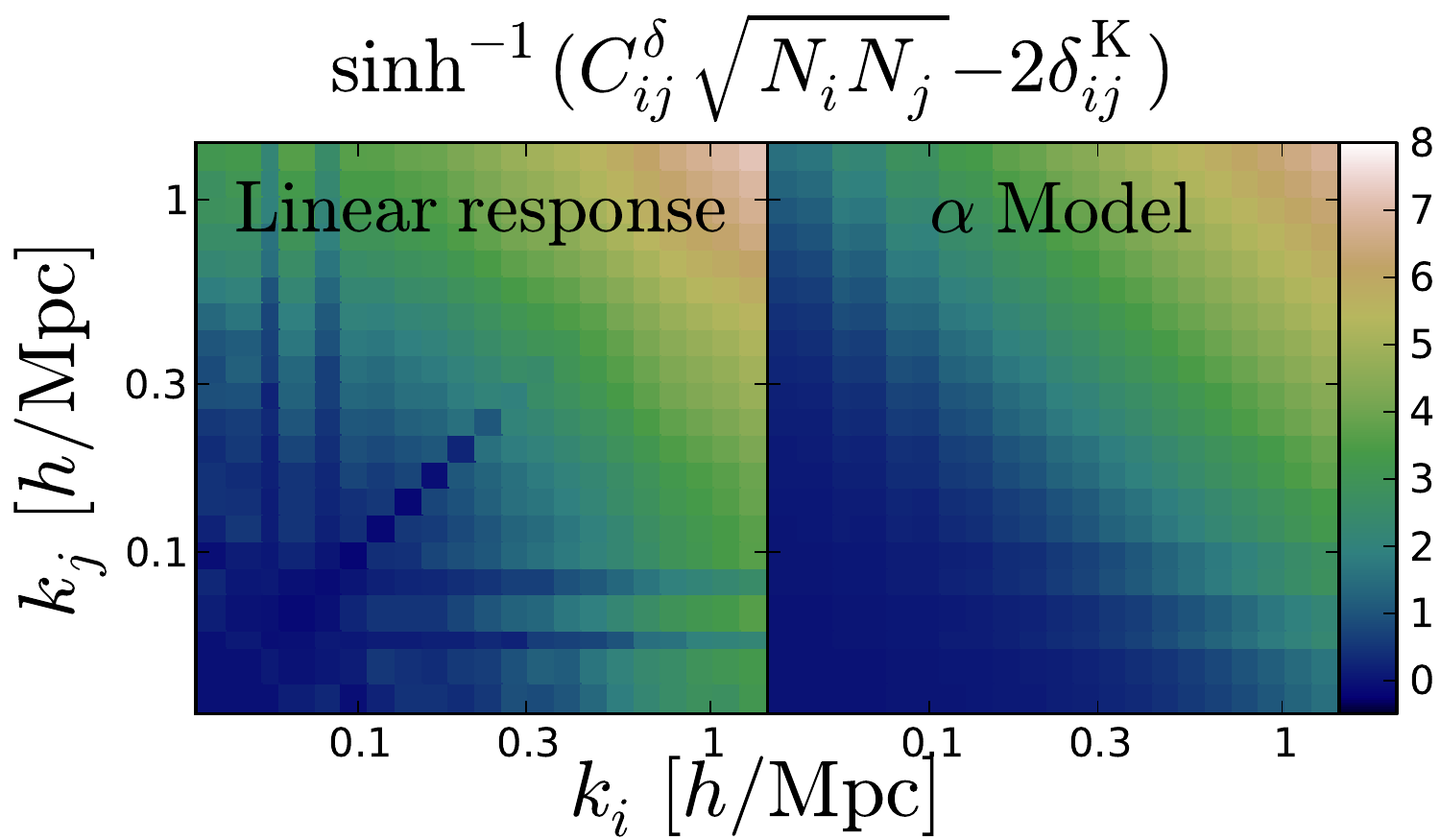}
  \end{center}
  \caption{The non-Gaussian part of the $\delta$ power-spectrum covariance $T_{ij}$, as measured from $G^\delta_{ij}$ using the linear-response model of Eq.\ (\ref{eqn:deltalin2}), and using an `$\alpha$ model' approximation to $T_{ij}$ found in \citep{Neyrinck2011a}.  Because of the steep increase, we use a $\sinh^{-1}$ transform for plotting, which becomes logarithmic for large values of its argument.  While the $\alpha$ model should not be taken too seriously as it is only an approximation, the qualitative agreement (except, perhaps, far from the diagonal) between the two plots suggests that the linear-response model captures most of the relevant effects.}
  \label{fig:compalpha}
\end{figure}

Fig.\ \ref{fig:compalpha} shows the non-Gaussian part of the $\delta$ power-spectrum covariance $T^\delta_{ij}\equiv C^\delta_{ij}(N_iN_j)^{1/2}-\delta^{\rm K}_{ij}$, both in the linear-response model from $G^\delta_{ij}$ in Eq.\ (\ref{eqn:deltalin2}), and from the fluctuating-multiplicative-bias model of \citet{Neyrinck2011a}, a rather accurate approximation to the covariance as measured from the Coyote Universe simulations \citep{LawrenceEtal2010}.  In this model, the non-Gaussian covariance is  given by $T^\delta_{ij} = \alpha (N_i N_j)^{1/2}$, where $\alpha$ is the fractional realization-to-realization variance of the nonlinear density-field variance in nonlinear-scale cells.   We use $\alpha=0.0035$ for this plot, which is $\alpha$ at $z=0$ as found by \citet{Neyrinck2011a}, scaled to the (256\hmpcnosp)$^3$ volume of the present simulations.

We emphasize that the $\alpha$ model is approximate, but qualitatively, it agrees with the linear-response model rather well, suggesting that additional terms in the covariance may indeed be subdominant.  The main discrepancies are in highly off-diagonal terms.

\section{Conclusions}
We used an $N$-body experiment to track where initial power-spectrum features get deposited in final-conditions density power spectra.  For the usual overdensity field $\delta$, our results qualitatively agree with the common wisdom that initial power migrates from large to small scales.  However, this seems to be largely because the $\delta$ field is dominated by overdense spikes.  When the density is transformed to have a more-Gaussian PDF, increasing the statistical weight of low-density regions (where patches imprinted with initial fluctuations expand rather than contract in comoving coordinates), initial spikes spread rather symmetrically, both upward and downward in scale.  In fact, in \precip, almost exclusively sensitive to underdense regions, initially small scales are magnified (AS13).  In these power spectra, initial small-scale spikes leave much more evidence at $z=0$ than in \pd.

The spread of power in the Gaussianized variables such as $\ln(1+\delta)$ is qualitatively captured by a toy model we give, in which patches imprinted with initial fluctuations expand or contract according to a spherical-collapse model.  In the future, it would be interesting to refine this model for greater accuracy, and investigate whether it might be modified successfully to other power spectra.

We also begin to apply our results to the theoretical question of how degrees of freedom present in the initial density field, essentially a sum of many spikes such as the ones we use, disappear from the final-conditions, coarse-grained density field.  However, this will require further measurements, because in our framework, fluctuations in the final power spectrum are a linear, invertible transformation of the initial power spectrum, given by a matrix $G_{ij}$.  In reality, though, information is lost because of a few neglected effects, which we will analyze in future work.  We do, however, find that $G_{ij}$ gives a rather accurate description of the \pd\ covariance matrix, suggesting that one of these effects (the nonlinear coupling of spike pairs) is not dominant.  Thus, `ringing' the initial power spectrum as we do offers an interesting technique to estimate power-spectrum covariances and Fisher information, by tracking the true, initial degrees of freedom in the Universe.

\section*{Acknowledgments}
We thank Nuala McCullagh, Julien Carron, Miguel Arag\'{o}n-Calvo, Xin Wang, Istv\'{a}n Szapudi and Alex Szalay for stimulating discussions.  This work used the Extreme Science and Engineering Discovery Environment (XSEDE), which is supported by NSF grant OCI-1053575.  We are grateful for support from NSF OIA grant CDI-1124403, and the Gordon and Betty Moore Foundation, and MCN is grateful for support from a New Frontiers in Astronomy and Cosmology grant from the Sir John Templeton Foundation. 

\bibliographystyle{hapj}
\bibliography{refs}

\end{document}